\documentclass{article}

\usepackage{amsmath, amssymb, graphicx, comment, amsthm, mathrsfs}
\usepackage{color}
\usepackage{hyperref}
\usepackage[makeroom]{cancel}
\newcommand{\be}{\begin{eqnarray}}
\newcommand{\ee}{\end{eqnarray}}
\newcommand{\nn}{\nonumber \\}

\def\bea {\begin{eqnarray}}
\def\eea {\end{eqnarray}}

\def\oo{&=&}
\def\nn{\nonumber\\}
\begin{document}
\today
\vspace{30pt}
\title{Dark Matter Spikes in Special K-Essence}

\begin{center}  
{ \bf \large Scalar Field Model for Dark Matter Spikes Surrounding Sgr A$^*$ and M87$^*$}

\vspace{1cm}

{Ramin G.~Daghigh$^{1}$, Michael D. Green$^{2}$, and Gabor Kunstatter$^3$,}
\end{center}

\centerline{\small \it $^1$ Natural Sciences Department, Metropolitan State University, Saint Paul, Minnesota, USA 55106}
\vskip 0 cm
\centerline{} 

\centerline{\small \it $^2$ Mathematics and Statistics Department, Metropolitan State University, Saint Paul, Minnesota, USA 55106}
\vskip 0 cm
\centerline{} 

\centerline{\small \it $^2$ Physics Department, University of Winnipeg, Winnipeg, MB Canada R3B 2E9}

\vspace{1cm}
\begin{abstract}
Theoretical models suggest that the adiabatic growth of a black hole immersed in dark matter can
lead to the formation of high density regions of dark matter, known as “spikes”, near the black hole event horizon. The density of these spikes is determined theoretically and observationally to be a power law of the form $\rho(r) \propto r^{-\gamma_{sp}}$. It has been shown that the spike can potentially have a detectable impact on the emitted gravitational waves and shadow radius of the central black hole.  In this work, we model the dark matter spike using a real scalar field with a non-standard potential. More specifically we ``reverse engineer'' the equations of motion to find a potential for the scalar field that permits a solution to the equations of motion with desired energy density and reasonable background geometry. We show that the emerging geometry is testable. In addition, the fact that the solution is derived from a covariant action makes it possible to study gravitational perturbations of the black hole in the presence of a spike including the backreaction of the spike.
\end{abstract}
\clearpage
\tableofcontents
\clearpage

\section{Introduction}

 Black holes are not isolated astrophysical objects.  Therefore, the environmental impact on the observational signals from black holes\cite{DMspikeGW1,DMspikeGW11,DMspikeGW2,DMspikeGW3,DMspikeGW4,DMspikeGW5,DMspikeGW6,DMspikeGW7,DMspikeGW66, DMspikeGW112, DMspikeGW1121} should be taken into account. In most galaxies, a dark matter (DM) halo is expected to surround the supermassive black hole at the galactic center.   It has been argued\cite{DMspikeformation1, GondoloSilk, DMspikeformation3} that the accretion of the halo into the central black hole will form a high-density region of DM, called a ``spike'', near the black hole event horizon.  A similar situation can potentially occur around intermediate-mass black holes\cite{DMspikeformation4,DMspikeformation5,DMspikeformation6}.  The spike is expected to start around twice the radius of the black hole event horizon\cite{Sadeghian} and its density drops following a power law\cite{GondoloSilk}. 

In order to explore the observational impact of the DM spike on the gravitational waves emitted from the central black hole or its  apparent shadow radius, one must first construct the spacetime metric.  This has been done analytically by several authors in \cite{BH-DMspike1, BH-DMspike2, DK-spike, DK-spike1}. The geometry of a black hole surrounded by a DM halo has also been studied by several authors\cite{DK-spike1, Halo-Jusufi1, Halo-Jusufi2, CardosoHalo, KonoplyaHalo}.  It has been shown\cite{DK-spike1} that the impact of the halo on the observational signals coming from the central black hole is negligibly small, but the spike can have a detectable impact.  For this reason, in this work, we focus mainly on the DM spike. 
 
In \cite{DK-spike, DK-spike1}, the spacetime metric was obtained analytically in closed form for a cold DM (CDM) spike/halo by solving the resulting Tolman-Oppenheimer-Volkoff (TOV) equations.  
The CDM was assumed to be a spherically symmetric anisotropic perfect fluid, and the anisotropic pressure in the radial direction was confirmed to be negligible. 
Next, \cite{DK-spike, DK-spike1} studied the ringdown waveform of a massless scalar field in the given fixed background metric.
This is the standard approach to understanding the ringdown associated with perturbations of a scalar field in a fixed curved background spacetime. This approach, however, does not take into account the backreaction of the DM spike on the background geometry, through which the perturbations are propagating. The DM in galactic halos is very sparse and gravitational effects weak so that one expects non-linear effects to be negligible. DM spikes, on the other hand, are substantially more dense near the horizon, where gravity is potentially strong and non-linear effects may be significant. 
The backreaction was studied for the analogous case of coupled electromagnetic and gravitational perturbations in the context of nonlinear electrodynamics in \cite{Zerilli, Moreno, Nomura, Okyay2021, Daghigh2021}. To accomplish this for a DM and gravity system, a phenomenological stress energy tensor for the spike is not sufficient. What is required is a set of coupled equations for the DM and the metric derived from a covariant action. These equations can yield not only the required background metric, spike density profile and pressure as a static solution, but  also the coupled dynamics of linearized perturbations to the metric and the DM.  The authors of \cite{ClusterSpike} used the Einstein cluster
framework to investigate the backreaction on the spacetime geometry of a DM spike, which is formed during the adiabatic growth of
a black hole embedded in a DM halo.  A similar methodology to \cite{ClusterSpike} applied to a DM halo and a DM spike, with zero radial pressure surrounding a black hole, can be found in \cite{EnviromImpact} and  \cite{TidalLove} respectively.  In this work, we take a different approach that involves a real scalar field. 

The widely used $\Lambda$CDM model describes DM that appears to make up over 20\% of the observed universe in terms of non-relativistic, non-interacting particles. While it is in principle possible to calculate backreaction in such a system, a field theory description is likely to be more manageable.  The $\Lambda$CDM model has had many successes over the years in explaining the broad features of the Universe that we see. However, it is not without its tensions\cite{TensionReview}.  Since DM, cold or otherwise, has yet to be detected directly, it is potentially useful to consider  possible scenarios other than the CDM. One such alternative that has been studied over the years  is Scalar Field Dark Matter(SFDM), which  has gained traction recently due to its potential to address some
of the issues with $\Lambda$CDM\cite{SFDM1, SFDM2, SFDM3}. As the name suggests, this model consists of a relativistic self-interacting scalar field and has been used to explain, among other things, the core-cusp problem associated with DM halos\cite{CuspProblem,Magnetic}, and the seeds of galaxy formation\cite{Seeds}. Various forms of scalar potential have been considered, including exponential\cite{ExpPot}, hyperbolic cosine\cite{Nunez}, and quadratic polynomial\cite{PolyDM}.\footnote{The SFDM goes by other names, including Fuzzy, Bose-Einstein or Wave DM. For a review of the phenomenology of SFDM, see \cite{WaveDM}} Among other alternative models for DM halos, there is also the k-essence, which uses non-canonical scalar fields\cite{Gauthier2010}.  The key difference between studying models for DM halos and DM spikes is that the latter occur in the strong field region near the horizon of the black hole.

In the present work, we attempt to gain information about the relevant form of the scalar field potential, $V(\phi)$, by using a form of reverse engineering. Specifically, we start with the proposed density profile of the DM spike near M87$^*$ and Sgr~A$^*$ and ask what form of scalar potential is required to yield the density as part of a static spherically symmetric solution of the Einstein equations that describes a reasonable spacetime metric. In other words, we consider a real scalar field action with potential that is designed to model the spike. Once the density of the spike is fixed as a function of the areal radius,  the TOV equations can be used to determine the mass function,  the radial and tangential pressures, and $g_{00}$ all as functions of the areal radius. The solutions depend on the three parameters that determine the density profile of the spike: the inner radius of the spike, the magnitude of the density at that radius, and the rate of exponential falloff. The outer radius is largely irrelevant to the quasinormal modes (QNMs) and ringdown waveform, which depend largely on the gravitational potential near the horizon. Once we have the metric components, pressure, and density as functions of $r$, it is straightforward to determine the scalar field $\phi(r)$ and its potential $V(r)$. By involution, this yields the scalar potential $V(\phi)$ for which our metric and scalar field are solutions. In the process, we show that zero radial pressure configurations are not possible, irrespective of the form of the potential. SFDM models of galactic halos with non-zero pressure were considered in \cite{Pressure1,Pressure2},  
while \cite{ClusterSpike} showed that the radial pressure cannot be zero in the DM spike modeled using the Einstein cluster framework. We also show analytically that at large distances the pressure must die off at the same rate or more slowly than the density.  Our numerical results favor the latter case with a slower drop in pressure.

The equations as described above are coupled ODE's and therefore difficult or impossible to solve analytically, so we initially proceed by finding numerical solutions. Remarkably, by using fairly standard fitting techniques we discover that there is a relatively simple analytic form for the potential, $V(\phi)$, that works for both Sgr A$^*$ and M87$^*$.  
If our action were a candidate for a fundamental theory of DM, the parameters in the scalar potential would be the same for both black holes but this turns out not to be the case. 
We emphasize, however, that the goal of this paper is not to construct a fundamental theory of DM, but to model the spike using a covariant action so that we can study the QNM spectrum and the ringdown of the coupled gravity-DM system. This will be done in a future publication. 

The remainder of the paper is organized as follows. In Sec.~\ref{sec: DMS}, we present the details of the DM spike profile formed due to the adiabatic growth of a black hole surrounded by a DM halo.  In Sec.~\ref{sec: Model}, we introduce the relativistic action of a real scalar field minimally coupled to gravity.  We then derive the TOV equations for a generic spherically symmetric spacetime that we can combine with the stress energy-tensor of the scalar field to find a differential equation that determines the radial pressure of the DM spike.  Once the pressure is determined numerically, we have all the ingredients to determine the scalar field potential.
In Sec.~\ref{sec: ScalarPotential}, we show that the scalar field potential can be fit almost perfectly using a Gaussian function.  In this section, we also determine the effective mass of the scalar field for our model.  In Sec.~\ref{sec:Results}, we present the observational consequences of our model by producing the ringdown waveform and calculating the shadow radius in the presence of a DM spike made of a real scalar field.  Finally, concluding remarks are given in Sec.~\ref{sec: conclude}.

\section{Dark Matter Spikes}
\label{sec: DMS}

Given a black hole with a mass $M_{\text{BH}}$ at a galactic center surrounded by a DM halo with an initial power law density profile
\begin{equation}
	\rho_{\text{DM}}(r)\simeq \rho_0 \left( \frac{r_0}{r} \right)^\gamma,
	\label{}
\end{equation}
where $\gamma$ is the power law index, with a range of $0 \le \gamma \le 2$, In this paper, we focus on the cases where $\gamma = 0$ and $\gamma = 1$.  In \cite{GondoloSilk}, it was shown within a primarily Newtonian approximation that during black hole formation in the presence of CDM a high density DM region, called a spike, will form adiabatically with a density profile\cite{DMspikeGW6, BH-DMspike2} 
\begin{equation}
	\rho_{\text{DM}}^\text{sp}(r)\simeq \rho_{\text{sp}} \left( \frac{R_\text{sp}}{r} \right)^{\gamma_\text{sp}}=\rho_\text{b} \left( \frac{r_\text{b}}{r}\right)^{\gamma_{\text{sp}}},
	\label{eq-SpikeDensity}
\end{equation}
where 
\begin{equation}
	\rho_{\text{sp}}= \rho_0 \left( \frac{R_\text{sp}}{r_0} \right)^{-\gamma},~R_{\text{sp}}=\alpha_\gamma r_0 \left( \frac{M_{\text{BH}}}{\rho_0 r_0^3}\right)^{\frac{1}{3-\gamma}}, ~ \text{and}~\gamma_{\text{sp}}=\frac{9-2\gamma}{4-\gamma}.
	\label{eq-spike-parameters}
\end{equation}
Here, $\rho_\text{sp}$ and $R_{\text{sp}}$ are the density and radius of the spike, respectively, at the outer edge.  Instead of  $\rho_\text{sp}$ and  $R_{\text{sp}}$, one can  use $\rho_{\text{b}}$ and  $r_{\text{b}}$, which are the density and radius of the spike at its inner edge. The general form of the spike was verified in an analysis by Sadeghian {\it et al.}\cite{Sadeghian} that took into account the relativistic nature of the gravitational effective potential. They found that the density vanished at  $r_\text{b}=2r_\text{BH}$, where
$r_\text{BH}=2GM_\text{BH}/c^2$ is the horizon radius of the central black hole, as opposed to $4r_\text{BH}$ as in the non-relativistic case\cite{GondoloSilk}. 
For relativistic CDM, zero density at $r_\text{b}=2r_\text{BH}$ is expected because it is the radius of the unstable circular orbit in the Schwarzschild geometry for a marginally bound particle. Any particle that manages to reach $r_b$ is necessarily captured by the black hole. 

Note that there are four parameters in the density profile: the inner and outer radii, the density at the inner (or outer) radius, and the exponent $\gamma_\text{sp}$. They differ for different galactic black holes. Some values of these parameters for M87$^*$ and Sgr A$^*$ that appear in the literature\cite{BH-DMspike2,DK-spike, M87data} are given in Table \ref{Table1}.

\section{The Model}
\label{sec: Model}
We wish to find a potential for a real Klein-Gordon scalar field such that  the static spherically symmetric solutions to the resulting Einstein-Klein-Gordon equations model closely the DM spikes in Sgr A and M87.
\subsection{Action}
The relativistic action of a real scalar field $\phi$ minimally coupled to gravity can be written in terms of the Einstein-Hilbert action $S_g$ as 
\bea
S &=& S_g+S_\phi
\label{eq:TotalAction}\\
S_g\oo \frac{c^3}{16\pi G}\int d^4x \sqrt{-g} \mathcal{R}
\label{eq:GravitationalAction}\\
S_\phi \oo \hbar \int d^4x \sqrt{-g}\mathscr{L}
\label{eq:ScalarAction}
\eea
where $\mathcal{R}$ is the Ricci scalar, $g$ is the determinant of the metric $g_{\mu \nu}$, and the Lagrangian density of the scalar field is given by
\bea
 \mathscr{L} =-\frac{1}{2} g^{\mu\nu}\partial_\mu \phi \partial_\nu\phi-V(\phi),
\label{eq:Lagrangian}
\eea
where $V(\phi)$ is the scalar field potential.  In the following, we scale our variables such that $G=c= r_\text{BH}=1$.  In this rescaled unit system, we can replace $\phi$ and $V(\phi)$ with unitless variables $\bar{\phi}=\phi \cdot r_\text{BH}$ and $\bar{V}(\phi)=V(\phi) \cdot r^4_\text{BH}$.  For simplicity, we omit the bar in the remainder of this paper.  

\subsection{Gravitational Equations}
We start with the most general spherically symmetric static metric in Schwarzschild-like coordinates
\bea
ds^2\oo -e^{\mu(r)}dt^2 + \frac{1}{f(r)}dr^2 +r^2d\Omega^{(2)},
\label{eq:metric}
\eea
where
\bea
f(r)\oo 1-\frac{2M(r)}{r}
\eea
with $M(r)$ being the combined mass of the central black hole and the DM spike within the radius $r$.
With the assumption of a perfect fluid stress tensor of the form
\bea
T^\mu_\nu = \hbox{diag}[-\rho(r), p_r(r), p_t(r), p_t(r)],
\label{eq:FluidEMT}  
\eea
 we arrive at the TOV equations\cite{Carroll}
 \bea
G_{tt}: \quad  \frac{dM(r)}{dr} \oo 4\pi r^2 \rho(r)
\label{eq:Gtt}\\
G_{rr}:\quad \frac{d\mu(r)}{dr} \oo 2\frac{M(r)+4\pi r^3 p_r(r)}{r\left[r-2M(r)\right]}
\label{eq:Grr}\\
\nabla_\nu T^{r\nu} = 0 \implies \frac{dp_r(r)}{dr}
\oo-[\rho(r)+p_r(r)] \frac{1}{2}\frac{d\mu(r)}{dr} + \frac{2}{r}(p_t-p_r)\nn
\oo-[\rho(r)+p_r(r)]\frac{M(r)+4\pi r^3p_r}{r\left[r-2M(r)\right]}+ \frac{2}{r}(p_t-p_r)\nn
\label{eq:MomentumConservation}
\eea
where $\rho(r)$, $p_r(r)$, and $p_t(r)$ are the mass density, radial pressure and transverse pressure respectively.
These equations were solved analytically in \cite{DK-spike, DK-spike1} to obtain the metric functions $M(r)$ and $\mu(r)$ for a power law spike density profile with negligible radial pressure to investigate the black hole ringdown and shadow. In the present work, however, we cannot neglect the pressure and will be forced to solve Eqs.~\eqref{eq:Grr} and \eqref{eq:MomentumConservation} numerically.  Equation \eqref{eq:Gtt} can be handled analytically for the spike density profile given in Eq.~\eqref{eq-SpikeDensity}, which gives
\bea
M(r)=M_\text{BH}+\frac{4 \pi \rho_{\text{sp}}}{3-\gamma_{\text{sp}}}\left[r^3 \left(\frac{R_{\text{sp}}}{r}\right)^{\gamma_{\text{sp}}}-  r_\text{b}^3 \left(\frac{R_{\text{sp}}}{r_\text{b}}\right)^{\gamma_{\text{sp}}} \right],
\label{eq-SpikeMass}
\eea
where the constant of integration is chosen so that $M(r_b)=M_\text{BH}$.


\subsection{Scalar Equation}
The scalar equation derived from (\ref{eq:ScalarAction}) by setting 
$\delta S_\phi / \delta \phi = 0$ is
\bea
\Box \phi - \frac{d V(\phi)}{d\phi} = 0.
\eea
In the case of spherical symmetry, with the static metric (\ref{eq:metric}) and static field $\phi(r)$, this becomes
\bea
\frac{1}{\sqrt{-g}}\left[\sqrt{-g}f(r)\phi'(r)\right]'-\frac{d V(\phi)}{d\phi} \oo 0,
\label{eq:ScalarEquation}
\eea
where $'$  represents the derivative with respect to the radial coordinate $r$ and
\bea
\sqrt{-g}=r^2 \sin(\theta) e^{\mu /2}(1-\tfrac{2M(r)}{r})^{-\frac{1}{2}}.
\eea
 The stress energy-tensor  obtained from the action (\ref{eq:ScalarAction}) is
 \bea
T_{\mu \nu}= -\frac{2}{\sqrt{-g}}\frac{\delta S_\phi}{\delta g^{\mu\nu}}=\hbar (\nabla_\mu \phi \nabla_\nu \phi + g_{\mu \nu}\mathscr{L}).
\label{eq:XnSpacelike}  
\eea
This is indeed of the form (\ref{eq:FluidEMT}). In the case of a static spherically symmetric field $\phi=\phi(r)$, the pressure and density can  be written as
\bea
\rho \oo -T^0{}_0 =   -\hbar \mathscr{L}  =  \hbar(X +V)  
\label{eq:EnergyDensity}\\
p_r \oo T^r{}_r = 
 \hbar(X  -V)\nn
\oo  \rho - 2 \hbar V
\label{eq:RadialPressure}\\
p_t \oo  T^\theta{}_\theta  = \hbar[g^{\theta\theta}( \partial_\theta\phi)^2 
+\mathscr{L}]=\hbar \mathscr{L}
\label{eq:TangentialPressure}
\eea
where $-X$ is the kinetic term $ -\frac{1}{2} g^{\mu\nu}\partial_\mu \phi \partial_\nu\phi$ that, under our assumptions, reduces to
\bea
X = \frac{1}{2} \left(1-\frac{2M(r)}{r}\right)(\phi')^2.
\label{eq:X}
\eea
 It will be important to check the energy conditions  for the system. They are (see for example \cite{PoissonBook}):
\bea
\rho\geq 0 \quad \& \quad\rho + p_i \geq 0; \qquad&&  \hbox{Weak Energy Condition}\\
\rho + p_i \geq 0; \qquad&& \hbox{Null Energy Condition}\\
 \rho+ \sum_i p_i\geq 0 \quad \& \quad \rho + p_i \geq 0; \qquad&& \hbox{Strong Energy Condition}\\
\rho \geq |p_i|; \qquad && \hbox{Dominant Energy Condition}
\eea
In the present case we find that\\
\bea
\rho \oo \hbar [X +V(\phi)]\\  
\rho + p_r \oo 2 \hbar X \\
\rho+ p_t \oo 0\\
\rho + p_r + 2 p_t \oo -2\hbar V(\phi) 
\eea
where $X$ is positive by definition (\ref{eq:X}). Thus, the Null Energy Condition is satisfied.  In the following procedure, we will specify the density $\rho$ of the spike to be positive and the potential $V(\phi)$ will turn out to be negative (see Section \ref{sec:Results}). This guarantees that the Weak and Strong Energy Conditions are satisfied.   
However, the Dominant Energy Condition is not satisfied when $V(\phi) < 0$ due to Eq.~\eqref{eq:RadialPressure}. In some cases, this violation can lead to superluminal propagation of the perturbations in the scalar field.  This is not the case for the real scalar field considered here. 


We note from Eqs.~(\ref{eq:EnergyDensity}, \ref{eq:RadialPressure}, \ref{eq:TangentialPressure}) that:
\bea
p_r-p_t = \rho + p_r
\eea
and use this to write the energy conservation law (\ref{eq:MomentumConservation}) as
\bea
\frac{dp_r(r)}{dr}
\oo -[\rho(r)+p_r(r)]\left(\frac{M(r)+4\pi r^3p_r}{r\left[r-2M(r)\right]}+\frac{2}{r}\right).\
\label{eq:MomentumConservation2}
\eea
Given the DM spike density profile \eqref{eq-SpikeDensity} and mass function \eqref{eq-SpikeMass}, we can solve for the radial pressure using the above differential equation.  We then can solve for $\mu(r)$ using Eq.~\eqref{eq:Grr}.  With the metric functions $M(r)$ and $\mu(r)$ in hand, we have the entire spacetime metric and can consequently determine the ringdown waveform and the shadow radius of a black hole surrounded by the scalar field DM spike.    Note that to solve for $p_r$, we need to choose an initial condition $p_r(r_b)$ that leads to a physically relevant asymptotic value  $p_r(\infty)=0$.  Numerically, it turns out that any initial condition at $r_b$ will lead to $p_r(\infty)=0$. 
However, we find that taking $p_r$ too small at $r_b$ leads to a negative pressure in some range of values of the radial coordinate and a negative value for the kinetic term $X=(\rho+p_r)/2$, which is not possible according to the definition of $X$ given in Eq.~\eqref{eq:X}. 

\begin{figure}[th!]
	\begin{center}	
        \includegraphics[height=3.5cm]{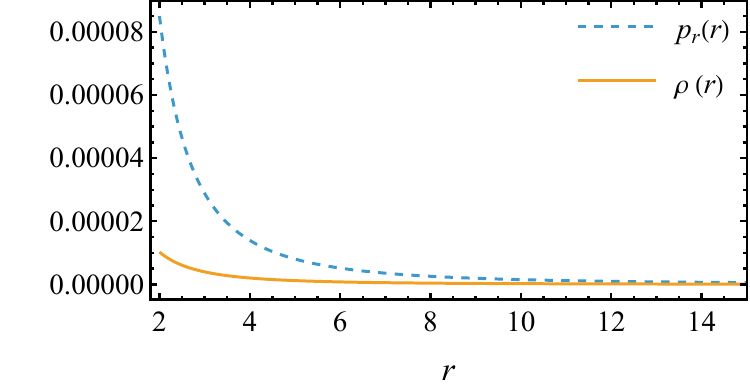}
         \includegraphics[height=3.6cm]{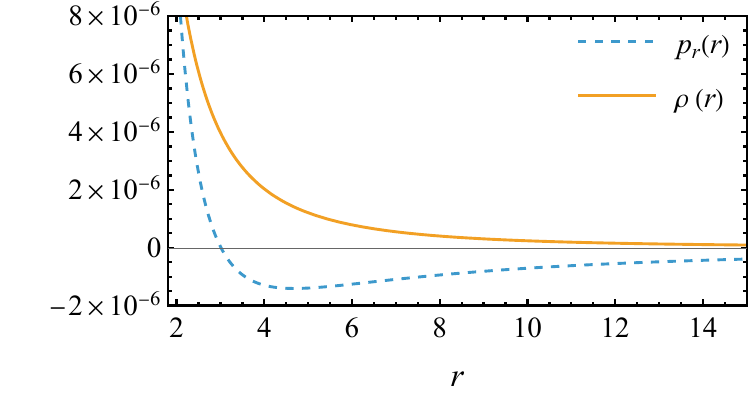}
	\end{center}
	\vspace{-0.7cm}
	\caption{\footnotesize The density and radial pressure of the DM spike as a function of radial coordinate for M87$^*$. We use the spike parameters in Table \ref{Table1} for $\alpha_\gamma=1.94$.  In the top panel, the initial condition is $p_r(r_b)=0.000085$, which is the minimum/critical value that leads to a positive pressure for all values of the radial coordinate.   For illustration, the behavior for a lower initial pressure, $p_r(r_b)=0.00001$, is shown in the lower panel.}
	\label{Fig: pressure}
\end{figure}

In Fig.~\ref{Fig: pressure}, we show the pressure profile for $M87$ with the minimum value of the initial condition $p_r(r_b)$ that leads to a positive pressure for all values of the radial coordinate.
We also show the pressure profile for an initial condition $p_r(r_b)$ less than the minimum/critical value.  For this lower initial pressure, for most values of $r$ the pressure is negative and $|p_r|>\rho$.  As mentioned above, this leads to $X<0$. This, in turn, causes a discontinuity in the first derivative of the metric and hence in the gravitational potential. However, this discontinuity is small and does not lead to observable effects. 
\begin{figure}[th!]
	\begin{center}	
        \includegraphics[height=4.5cm]{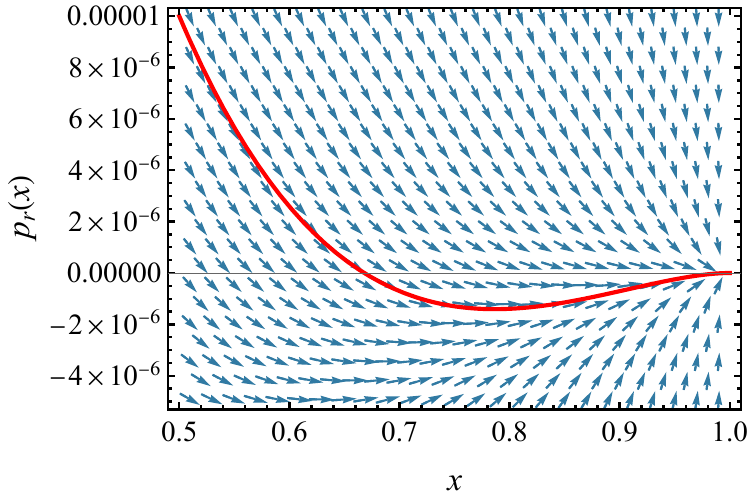}
	\end{center}
	\vspace{-0.7cm}
	\caption{\footnotesize The slope field of the differential Eq.~\eqref{eq:MomentumConservation2} for radial pressure of the DM spike as a function of $x=1-2 M_0/r$ for M87$^*$ together with the specific case where $p_r(r_b)=0.00001$. We use the spike parameters in Table \ref{Table1} for $\alpha_\gamma=1.94$.  The slope field clearly shows that the radial pressure has to go to zero as $r\rightarrow \infty$.}
	\label{Fig: pressurefield}
\end{figure}

\subsection{Asymptotic form of the pressure}

We can analytically explore the asymptotic behavior of $p_r$ as $r \rightarrow \infty$.  First, we note that since $\rho\propto r^{-\gamma_{\hbox{sp}}}$ with $2<\gamma_{\hbox{sp}}<3$, $M(r)$ asymptotically drops proportional to $r^{3-\gamma_\text{sp}}$ and can be neglected both in the numerator and denominator on the right hand side of (\ref{eq:MomentumConservation2}).  This leads to
\bea
\frac{dp_r(r)}{dr}
\oo -[\rho(r)+p_r(r)]\left(4\pi r p_r(r)+\frac{2}{r}\right).\
\label{eq:MomentumConservation3}
\eea
We now have three options.  First, assume $p_r$ asymptotically drops faster than $\rho$.  In this case, Eq.~\eqref{eq:MomentumConservation3} reduces to
\bea
 \frac{dp_r(r)}{dr}
\oo -\frac{2 \rho(r)}{r},\
\label{}
\eea
which leads to $p_r(r) \propto r^{-\gamma_\text{SP}}$.  This contradicts our initial assumption that $p_r$ drops faster than $\rho$. 
Next we assume that the radial pressure drops off at the same rate as the density, i.e.~$p_r(r) \sim A \rho(r)$ for some constant $A$.  Then 
\bea
 \frac{dp_r(r)}{dr}
\oo -\frac{2}{r}\left[\rho(r)+p_r(r)\right].\
\label{}
\eea
This leads to $p_r(r) \sim \tfrac{2}{\gamma_\text{SP}-2}\rho(r)$, which is consistent with our assumption. However, our numerical calculations suggest that the radial pressure does not drop off at the same rate as the density.

Finally, we assume that the pressure goes to zero more slowly than the density as $r \rightarrow \infty$.  In this case
\bea
 \frac{dp_r(r)}{dr}
\oo -4\pi r p_r^2(r)-\frac{2 p_r(r)}{r}
\label{eq: two-terms}
\eea
which has the solution
\bea
 p_r(r) \propto \frac{1}{ r^2 \ln{r}}.
\label{eq:prLog}
\eea
This is consistent with the initial assumption that $p_r$ drops slower than $\rho$. In contrast to the previous case, the form (\ref{eq:prLog}) is supported by our numerical calculations.  Therefore, in the next section, we use a model for the radial pressure that exhibits this asymptotic behavior. 



\begin{figure}[th!]
	\begin{center}
		\includegraphics[height=4.cm]{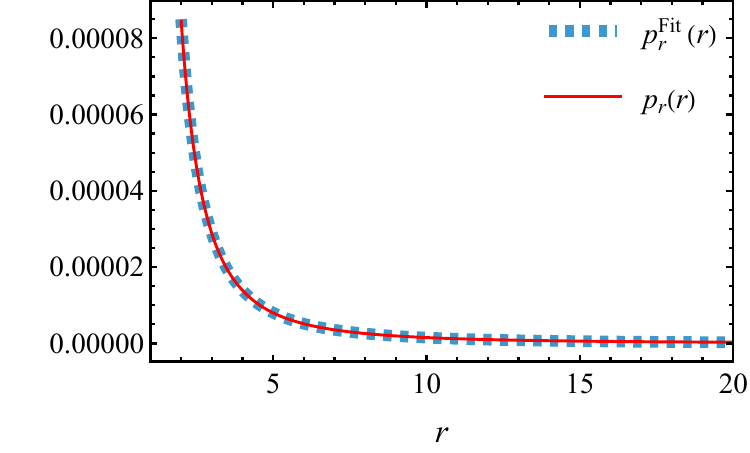}
        \hspace*{0.4cm}
        \includegraphics[height=4.cm]{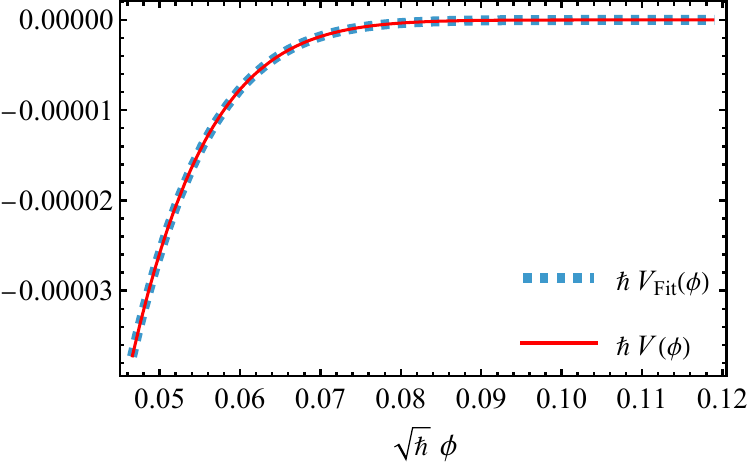}
        \hspace*{0.4cm}
        \includegraphics[height=4.cm]{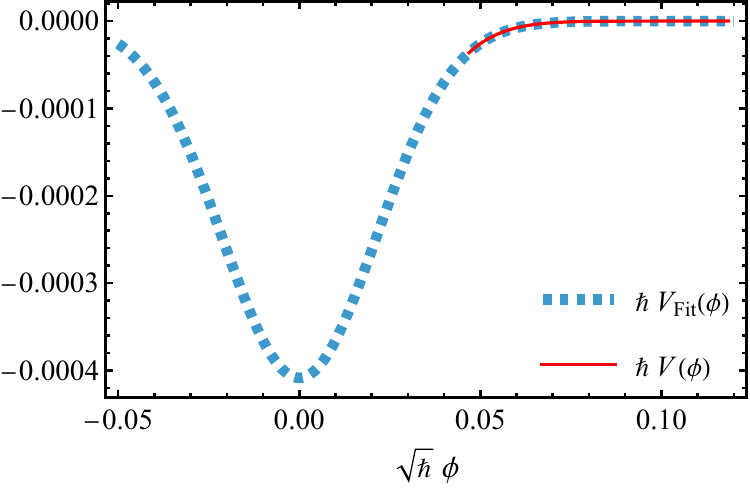} 
        \hspace*{0.8cm}
        \includegraphics[height=4.1cm]{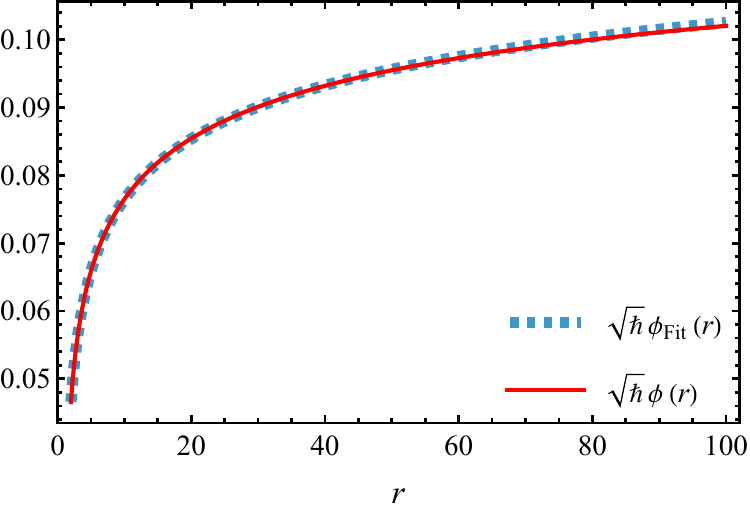}
	\end{center}
	\vspace{-0.7cm}
	\caption{\footnotesize The top plot shows the radial pressure $p_r$ as a function of $r$, in solid red, compared to the fitted function \eqref{eq:Pressure-fit} in dashed blue.  The second plot shows the scalar potential $V$ as a function of $\phi$ and the fitted Gaussian function \eqref{eq:Vphi-fit} over the interval $r\in [r_b, 1000]$.  The third plot is the expanded version of the second plot with the minimum of the scalar potential included in the plot. As a consistency check for our fitting method, in the fourth plot, we show how well the scalar field $\phi$ as a function of $r$ is fitted using the analytic pressure function \eqref{eq:Pressure-fit}.  To generate these plots, we use the spike parameters for M87$^*$ in Table \ref{Table1} for $\alpha_\gamma=1.94$.}
	\label{Fig: ScalarPotential}
\end{figure}

\section{Determination of Scalar Potential for M87}
\label{sec: ScalarPotential}

In this section, we find a simple function $V(\phi)$ to fit the scalar potential so that we can estimate the mass of the scalar field.  
We first find a model for the radial pressure consistent with the asymptotic behavior in Eq.~\eqref{eq:prLog}, namely
\bea
p_r(r)=\frac{1}{r^2(a_p+b_p \ln r)}.
\label{eq:Pressure-fit}
\eea
Fitting this to the pressure profile shown in the top panel of Fig.~\ref{Fig: pressure} we find 
$a_p=1401.8$ and $b_p=2233.1$.

Knowing $p_r(r)$  and $\rho(r)$ we can determine $V(r)$ and $X(r)$ from Eqs.~\eqref{eq:EnergyDensity} and (\ref{eq:RadialPressure}). Using Eq.~\eqref{eq:X}, $\phi(r)$ is then obtained from $X(r)$ by numerical integration. Finally, $V(\phi)$ is determined by involution. 
The resulting $V(\phi)$ can be fit almost perfectly using a Gaussian function of the form
\bea
\hbar V(\phi)=- A e^{-\frac{(\sqrt{\hbar}\phi -C)^2}{B^2}},
\label{eq:Vphi-fit}
\eea
with the numerical values $A=0.000408383$ and $B=0.0301035$ used in the second and third plots in Fig.~\ref{Fig: ScalarPotential}.  The constant $A$ has dimensions $[L^{-2}]$, while $B$ is dimensionless. $C$ can be set  to zero without loss of generality by an appropriate choice of the constant of integration in $\phi$. We  have therefore shown that the DM spike can be modeled by a remarkably simple two parameter real Klein-Gordon scalar action with Gaussian potential.

From Eq.~\eqref{eq:RadialPressure}, we know 
\bea
\hbar V(r)=\frac{\rho(r)-p_r(r)}{2},
\label{eq:Vr-fit}
\eea
which can now be written in analytic form using the DM spike density profile Eq.~\eqref{eq-SpikeDensity} for $\rho(r)$ and the fitted pressure function given in Eq.~\eqref{eq:Pressure-fit}.
As a consistency check, we can now find an analytic function for $\phi(r)$ by combining Eqs.~\eqref{eq-SpikeDensity}, \eqref{eq:Pressure-fit}, \eqref{eq:Vphi-fit}, and \eqref{eq:Vr-fit}.  As shown in the last plot of Fig.~\ref{Fig: ScalarPotential}, this analytic function matches well with the numerically obtained data for $\phi(r)$.  This further supports the validity of our fitting method.

Using the above model, the effective mass of the scalar field can be computed via
\bea
\frac{m^2_\text{eff} c^4}{\hbar^2 c^2} =\frac{d^2 V(\phi)}{d\phi^2}\Big|_{\phi=0} \cdot  r_\text{BH}^{-2}= \frac{2A}{B^2}\cdot  r_\text{BH}^{-2}= 0.9013 ~ r_\text{BH}^{-2},
\eea
where we have used the values given in Table \ref{Table1} for M87$^*$ with $\alpha_\gamma=1.94$.
Note that the mass of the scalar field depends on  $r_\text{BH}$ and consequently on the mass of the central black hole. This is not surprising since it is the only dimensional parameter in the problem and sets the scale. However, it is a potential weakness of the model that prevents it from being considered as a fundamental description of dark matter in general.  In particular, the calculations in this section are done for M87.  If we do the same calculations,  using the values given in Table \ref{Table1} for Sgr A$^*$ with $\gamma_\text{sp}=7/3$, we find the effective mass of the scalar field to be
\bea
\frac{m^2_\text{eff} c^4}{\hbar^2 c^2} =\frac{d^2 V(\phi)}{d\phi^2}\Big|_{\phi=0} \cdot  r_\text{BH}^{-2}= 0.9533 ~ r_\text{BH}^{-2},
\eea
which ends up three orders of magnitude larger than the mass of the scalar field around M87.  More specifically, we find the scalar filed mass to be $m_\text{eff} c^2 \approx 6.56\times 10^{-20}$ eV for  M87 and $m_\text{eff} c^2 \approx 1.03\times 10^{-16}$ eV for Sgr A.

It must be noted that the calculation of the mass depends on the properties of $V(\phi)$ around $\phi=0$, which is far from the calculated data seen in the third panel of Fig. \ref{Fig: ScalarPotential}.  Therefore, this should be taken with a grain of salt since it is extremely dependent on the particulars of our model.

\section{Observational Consequences: Ringdown and Shadow}
\label{sec:Results}

In this section, we look at the observational consequences of the presence of a DM spike made of a real scalar field around M87$^*$ and Sgr A$^*$.  More specifically, we numerically produce the ringdown waveform for the scalar field perturbations and calculate the enlargement of the black hole shadow radius in the presence of the spike.  Note that our ringdown calculations, where we look at the scalar field perturbations in the fixed  spacetime background of a black hole surrounded by a spike, are qualitative in nature.  For observationally relevant ringdown waveforms, one has to look at the coupled scalar and gravitational perturbations analogues to coupled electromagnetic and gravitational perturbations in Reissner-Nordstr\"om black holes derived in \cite{Zerilli}.

\subsection{Impact on the Ringdown Waveform}
The study of black
hole perturbations can be simplified by exploring the master wave equation 
\bea
\frac{\partial^2\Psi}{\partial t^2}+\left(-\frac{\partial^2}{\partial r_*^2}+\mathcal{V}(r)\right)\Psi=0.
\label{WE-time}
\eea
In the above equation, $r_*$ is the tortoise coordinate linked to the radial coordinate according to
\bea
dr_*=\frac{dr}{\sqrt{e^{\mu(r)}f(r)}},
\label{tortoise}
\eea
for a spherically symmetric metric given in Eq.~\eqref{eq:metric},
and 
\bea
\mathcal{V}(r)= e^{\mu(r)}\frac{l(l+1)}{r^2}+ \frac{1}{2r} \frac{d}{dr} \left[ e^{\mu(r)}f(r) \right]
\label{eq-scalarV}
\eea
is the Regge-Wheeler effective potential for scalar field perturbations.  Since the fundamental QNM of geometric perturbations in a black hole spacetime has the multipole number $l=2$, in the rest of the paper, we will focus on scalar perturbations with $l=2$.

In Table \ref{Table1}, we provide some of the data obtained from the literature  for the DM spike surrounding the supermassive black holes Sgr A$^*$ and M87$^*$.  In the DM made of a scalar field, in which the radial pressure $p_r$ cannot be ignored, the redshift factor, $\mathcal{C}$, is significantly larger than the isotropic/anisotropic DM spike, studied in \cite{DK-spike, DK-spike1}, where $p_r= 0$.  For more details on the redshift factor, see \cite{DK-spike1}.

\begin{figure}[th!]
	\begin{center}	
        \includegraphics[height=5.cm]{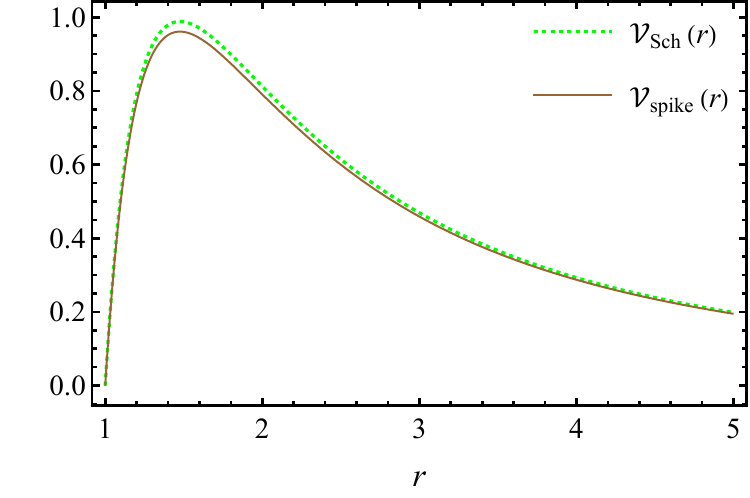}
	\end{center}
	\vspace{-0.7cm}
	\caption{\footnotesize The effective potential for the M87 black hole, in solid brown, surrounded by a DM Spike made of a scalar field.  
    For comparison, we also show the Regge-Wheeler potential for the Schwarzschild case in dotted green.  
    We use the spike parameters in Table \ref{Table1} for $\alpha_\gamma=1.94$.}
	\label{Fig: RW-ds}
\end{figure}

In Fig.~\ref{Fig: RW-ds}, we show the effective potential for M87$^*$ surrounded by a DM spike made of a scalar field.  We choose the case in Table \ref{Table1}, where $\alpha_\gamma=1.94$.  This case has the largest redshift factor.  For comparison, we also show the Regge-Wheeler potential for the Schwarzschild black hole with no spike. The impact of the DM spike made of a scalar field has a visible impact on the shape of the potential. 

\footnotesize
\begin{table}[htbp]
\centering
\caption{DM Spike surrounding {Sgr A$^*$} and {M87$^*$}  supermassive black holes ($\mathcal{C}$ is the redshift factor).}
\begin{tabular}{ccccccc}  	
	\hline
	\vspace{-0.3cm}\\
	$\gamma_\text{sp}$ & $M_\text{BH}$ ($10^6 M_\odot$)& $\alpha_\gamma$ & $R_{\text{sp}}$ (kpc) &  $\rho_{\text{sp}}$ (g cm$^{-3}$)  & $\mathcal{C}$  & $\mathcal{C}$ \\ 
     & & &   &  &  ($p_r=0$) &\\ 
	\hline 
    	$7/3$& $4.1 ~(\text{Sgr A$^*$\cite{BH-DMspike2}})$  & $1.94$ & $0.235$ &  $2.89\times 10^{8}$  &  1.0002  &  1.002\\
        $9/4$ & $4.1 ~(\text{Sgr A$^*$\cite{BH-DMspike2}})$ & $1.94$ & $0.910$ &   $2.59\times 10^{8}$   & 1.00002  &  1.0001 \\
	$7/3$& $6400~(\text{M87$^*$\cite{M87data}})$  & $0.1$ & $0.219$ &  $1.21\times 10^{9}$ & 1.00006 &  1.0009\\ 
	$7/3$& $6400 ~(\text{M87$^*$\cite{DK-spike}})$ & $1.94$ & $4.26$ &   $4.54\times 10^{11}$   & 1.004 &1.0285\\
    \end{tabular}
    \label{Table1}
    \end{table}

\normalsize

\begin{figure}[th!]
	\begin{center}	
        \includegraphics[height=5.cm]{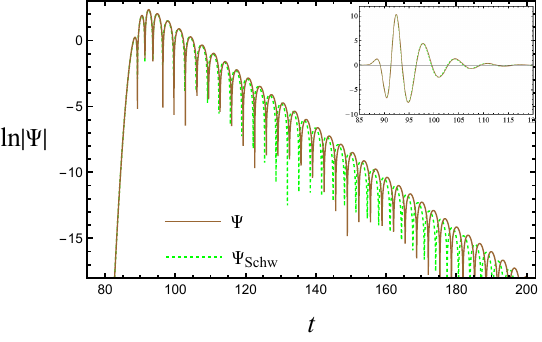}
	\end{center}
	\vspace{-0.7cm}
	\caption{\footnotesize The inset shows the ringdown for the M87 black hole, in solid brown, surrounded by a DM spike made of a scalar field.  The ringdown results from a perturbation emanating from just outside the event horizon moving outward.  For comparison, we also show the ringdown waveform for the Schwarzschild case in dotted green.  The main plot shows the logarithm of the ringdown waveform, where the oscillations are more pronounced.  We use the spike parameters in Table \ref{Table1} for $\alpha_\gamma=1.94$.  }
	\label{Fig: RW-ds-ringdown}
\end{figure}

\begin{figure}[th!]
	\begin{center}	
        \includegraphics[height=5.cm]{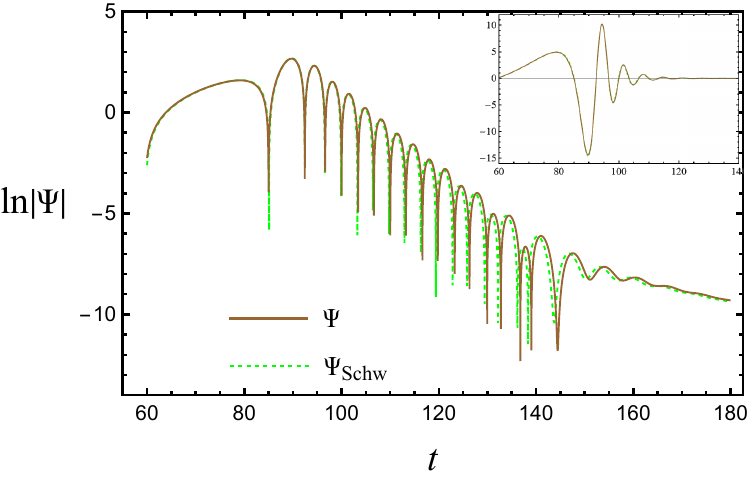}
	\end{center}
	\vspace{-0.7cm}
	\caption{\footnotesize The inset shows the ringdown for the M87 black hole, in solid brown, surrounded by a DM spike made of a scalar field.  The ringdown results from a perturbation emanating from far outside the black hole moving inward.  For comparison, we also show the ringdown waveform for the Schwarzschild case in dotted green.  The main plot shows the logarithm of the ringdown waveform, where the oscillations are more pronounced.  We use the spike parameters in Table \ref{Table1} for $\alpha_\gamma=1.94$.  }
	\label{Fig: RW-ds-ringdown-inward}
\end{figure}

In Fig.~\ref{Fig: RW-ds-ringdown}, we generate the ringdown waveform for the potential shown in Fig.~\ref{Fig: RW-ds}.  For comparison, we also show the ringdown waveform for the Schwarzschild case.  The difference is visible.  To obtain the ringdown waveform, we use the finite difference method to numerically solve the time-dependent wave equation (\ref{WE-time}) using the initial data
\bea
\Psi(r_*,0)={\cal A} \exp \left(- \frac{(r_*-\bar{r}_{*})^2}{2\sigma^2} \right),~  \partial_t \Psi|_{t=0}=-\partial_{r_*} \Psi(r_*, 0)~,
\label{GaussianWave}
\eea 
where we use $\sigma=1 ~r_\text{BH}$, $\bar{r}_*=-40 ~r_\text{BH}$, and ${\cal A}=30 ~r_\text{BH}^{-2}$.  We choose the observer to be located at $r_*=50 ~r_\text{BH}$. In all the cases studied here,  the height of the effective potential at $r_*=50 ~r_\text{BH}$, which is inside the spike region, is small ($\lessapprox 3\times10^{-3} r_\text{BH}^{-2}$) compared to the peak. Therefore, we do not expect a significant difference in the results if the observer is further away.  In Fig.~\ref{Fig: RW-ds-ringdown-inward},  we show the ringdown waveform for the same potential shown in Fig.~\ref{Fig: RW-ds}, but this time the initial Gaussian wavepacket emanates from far outside the black hole ($\bar{r}_*=40 ~r_\text{BH}$) moving inward.

We then can extract the first two  QNM frequencies ($\omega_0$ and $\omega_1$) from the ringdown waveforms shown in Figures \ref{Fig: RW-ds-ringdown} and \ref{Fig: RW-ds-ringdown-inward} using the Prony method\cite{Prony}, a numerical procedure that fits N data points by as many purely damped exponentials as necessary. To test the Prony method, we first calculate the QNM of scalar perturbations for the Schwarzschild case for $l=2$.  The result is $\omega_0=0.9673072-0.1935305i$ and $\omega_1=0.9277136-0.5912752 i$ for the outgoing initial data in Eq.\eqref{GaussianWave}.  For the ingoing initial data, we find $\omega_0=0.9673072-0.1935305i$ and $\omega_1=0.9277404-0.5910854 i$.  These results agree well with $\omega_0=0.967290 -0.193518 i$ and $\omega_1=0.927701 -0.591208 i$, which are found using the continued fraction method\cite{DGMK}.  For the ringdown waveform presented in Figure \ref{Fig: RW-ds-ringdown}, generated by an outgoing initial pulse, we find $\omega_0=0.9537246-0.1902302i$ and $\omega_1=0.9155173-0.5804792 i$.
For the ringdown waveform generated by an ingoing pulse, shown in Figure \ref{Fig: RW-ds-ringdown-inward}, we find $\omega_0=0.9537246-0.1902302i$ and  $\omega_1=0.9155689-0.5803163 i$. The results for both ingoing and outgoing initial data are consistent, and they differ up to the second decimal place from the Schwarzschild result. 

To handle the discontinuity in the potential at $r_*^b = r_*(r_b)$, for each $t$-value we first calculate values for $\Psi_-(r_*)$ and $\Psi_+(r_*)$ to the left and right, respectively, of $r_*^b$. At $r_*^b$ we require $\Psi$ to satisfy the junction conditions $\Psi_-(r_*^b)=\Psi_+(r_*^b)$ and $\partial_{r_*} \Psi_-(r_*^b) = \partial_{r_*} \Psi_+(r_*^b)$.  The value of $\Psi(r_*^b)$ is then obtained by using a second-order, one-sided approximation of the $r_*$-derivatives and solving the resulting equation for $\Psi(r_*^b)$.  For more details, see \cite{CF-discontinuity}.

\subsection{Black Hole Shadow Radius Enlargement}

It has been shown in \cite{DK-spike1} that the shadow radius of a Schwarzschild black hole surrounded by DM can be calculated using a simple formula
\begin{equation}
	r_\text{sh}^\text{sp} \sim    r_\text{ph} \sqrt{\frac{1}{A(r_\text{ph})/\mathcal{C}} }=3\sqrt{3\mathcal{C}}M_\text{BH} ,
\label{}
\end{equation}
where $\mathcal{C}$ is the redshift factor due to the presence of the DM spike.  Therefore, the black hole shadow radius will appear to be $\sqrt{\mathcal{C}}$ times larger due to the presence of DM.   For the shadow radius of M87$^*$, the case of $\alpha_\gamma=1.94$ in Table \ref{Table2} produces an enlargement of $1.4\%$ in the shadow radius.  This enlargement is accessible with the current Event Horizon Telescope data\cite{EHT-M87-PRIMO}. On the other hand, the case of $\alpha_\gamma=0.1$ in Table \ref{Table2} produces an enlargement of $0.0045\%$ in the shadow radius, which is roughly an order of magnitude below the sensitivity of the current Event Horizon Telescope data\cite{EHT-M87-PRIMO}.  In the case of Sgr A$^*$, the best case scenario for the expected spike parameters is an order of magnitude below the sensitivity of the Event Horizon Telescope.
\footnotesize
\begin{table}[htbp]
\centering
\caption{Shadow radius of {Sgr A$^*$} and {M87$^*$}  in the presence of DM}
\begin{tabular}{ccccc}  	
	\hline
	\vspace{-0.3cm}\\
	$\gamma_\text{sp}$ & $M_\text{BH}$ ($10^6 M_\odot$) & $\alpha_\gamma$  & $M_{\text{total}}^{\text{sp}}$ ($M_\odot$)  & radius enlargement\\ 
	\hline 
    	$7/3$& $4.1 ~(\text{Sgr A$^*$\cite{BH-DMspike2}})$  & $1.94$ &    $8.00 \times 10^{-23}$  &  0.1\%\\
        $9/4$ & $4.1 ~(\text{Sgr A$^*$\cite{BH-DMspike2}})$ & $1.94$ &    $1.39 \times 10^{-24}$  &  0.005\% \\
	$7/3$& $6400~(\text{M87$^*$\cite{M87data}})$  & $0.1$ & $4.10 \times 10^{-22}$  &  0.045\%\\ 
	$7/3$& $6400 ~(\text{M87$^*$\cite{DK-spike}})$ & $1.94$ & $2.12 \times 10^{-23}$  & 1.4\%\\
    \end{tabular}
    \label{Table2}
    \end{table}
\normalsize

\section{Conclusion}
\label{sec: conclude}

In this paper, we model the DM spike near M87$^*$ and Sgr A$^*$ using a real scalar field.  By combining the proposed density profile of the DM spike with the TOV equations and the stress-energy tensor of the scalar field, we were able to reverse engineer the scalar field potential, $V(\phi)$. The calculations are done numerically.  However, the form of the scalar field potential fits well with a simple Gaussian function given in Eq.~\eqref{eq:Vphi-fit}.

In this model, the pressure cannot be ignored.  
This agrees with the results of \cite{ClusterSpike}, which shows the impact of the radial pressure on the spacetime geometry cannot be neglected in the Einstein cluster model.
The presence of pressure increases the redshift factor compared to the isotropic/anisotropic DM spike with zero radial pressure studied in \cite{DK-spike, DK-spike1}.  This increased redshift leads to an enlargement in the shadow radius of M87$^*$, which  is detectable for the more optimistic case in Table \ref{Table2} with the current Event Horizon Telescope data\cite{EHT-M87-PRIMO}.  For the other cases , the shadow radius enlargement is at least an order of magnitude below the sensitivity of the current data. 

Using qualitative analysis of the scalar field perturbations in a fixed spacetime background, we also show that the presence of the scalar filed pressure increases the impact of the spike on the ringdown waveform.  More specifically, we find that QNMs in the presence of the spike differ up to the
second decimal place from the Schwarzschild QNMs.

The availability of the scalar field potential $V(\phi)$ and its covariant action makes it possible to include the backreaction of the DM spike on the gravitational perturbations of the central black hole, which has been ignored in the earlier literature\cite{GravWaveSpike, GravWaveSpike1, GravWaveSpike2, GravWaveSpike3}.
\\[10pt]
\noindent 
{\bf Acknowledgements}\\[5pt]
GK gratefully acknowledges that this research was supported in part by Discovery Grant number  2018-04090 from the Natural Sciences and Engineering Research Council of Canada.

\end{document}